\documentclass{article}
\pdfoutput=1

\usepackage{arxiv}

\usepackage[utf8]{inputenc} 
\usepackage[T1]{fontenc}    
\usepackage{hyperref}       
\usepackage{url}            
\usepackage{booktabs}       
\usepackage{amsfonts}       
\usepackage{nicefrac}       
\usepackage{microtype}      
\usepackage{lipsum}
\usepackage{siunitx}
\usepackage[linesnumbered,ruled,vlined]{algorithm2e}
\usepackage{float}
\usepackage{amsmath}
\usepackage{enumerate}
\usepackage{cite}
\usepackage{xcolor}
\usepackage{graphicx}
\usepackage{braket}
\usepackage{comment}
\usepackage{caption}
\usepackage{cleveref}
\graphicspath{ {./images/} }
\newcommand{\tr}{\mathrm{tr}}

\title{Compare Similarities Between DNA Sequences Using Permutation-Invariant Quantum Kernel}

\author{
 Chenyu Shi\\
  Applied Quantum Algorithms Leiden, Leiden University\\
  Leiden Institute of Advanced Computer Science, Leiden University\\
  \texttt{c.shi@liacs.leidenuniv.nl}\\
  \And
 Gabriele Leoni\\
  Joint Research Centre, Directorate F Health and Food, Digital Health Unit\\
  European Commission\\
  \texttt{gabriele.leoni@ec.europa.eu}\\
  \And
  Mauro Petrillo\\
   Seidor\\
   Seidor Italy S.r.l., Milan, Italy\\
  \texttt{mauro.petrillo@seidor.com}\\
  \And
   Antonio Puertas Gallardo \footnotemark[1] \\
   Joint Research Centre, Directorate F Health and Food, Digital Health Unit\\
   European Commission\\
   \texttt{antonio.puertas-gallardo@ec.europa.eu}\\
  \And
   Hao Wang \footnotemark[1]\\
  Applied Quantum Algorithms Leiden, Leiden University\\
  Leiden Institute of Advanced Computer Science, Leiden University\\
  \texttt{h.wang@liacs.leidenuniv.nl}\\
}

\begin{document}

\maketitle
\footnotetext[1]{Corresponding Authors}

\begin{abstract}
Computing the similarity between two DNA sequences is of vital importance in bioscience, yet it can be computationally expensive on classical hardware. For example, the edit distance with move operations (EDM), a DNA similarity measure of interest in biology, is proven to be NP-Complete to compute exactly on classical hardware. Recently, applied quantum algorithms have been anticipated to offer potential advantages over classical approaches. In this paper, we propose a novel variational quantum kernel model served as a surrogate model for estimating similarity between DNA sequences defined by EDM.
Since the EDM metric exhibits a pairwise permutation-insensitive property, we incorporate a permutation-invariant structure into the variational quantum kernel to approximate this symmetry. Furthermore, to encode the four nucleotide bases as quantum states, we introduce a theoretically motivated encoding scheme based on symmetric informationally complete positive operator-valued measure (SIC-POVM) states. This encoding ensures mutual equivalence among bases, as each pair of symbols is mapped to quantum states that are equidistant on the Bloch sphere. We experimentally show that, equipped with the permutation-invariant circuit design and mutual-equivalence encoding, the proposed quantum kernel model achieves strong performance in approximating the similarity defined by EDM. Compared with classical kernel learning methods, our quantum approach achieves significantly higher accuracy while using substantially fewer trainable parameters.
\end{abstract}

\section{Introduction}
Comparing the similarity between DNA sequences is a fundamental task in bioinformatics and comparative genomics \cite{heather2016sequence}. With the rapid development in bioscience techniques, the need for efficient and accurate DNA sequence similarity comparison has become increasingly important, particularly in applications such as antimicrobial resistance (AMR) gene detection \cite{feldgarden2021amrfinderplus, galhano2021antimicrobial}. Numerous classical methods have been proposed for evaluating DNA sequence similarity. For example, the Needleman-Wunsch algorithm utilizes edit distance for global sequence alignment \cite{needleman1970general}. Additionally, FASTA \cite{pearson1988improved} and BLAST \cite{altschul1990basic} employ heuristic approaches to compute similarity, both of which have achieved significant success.

However, due to the large scale of DNA sequence data, many edit distance-based classical methods can be computationally expensive and resource-consuming. For example, the edit distance with move operations (EDM), which is of great interest in biological sequence analysis, has been proven to be NP-Complete to compute exactly on classical hardware \cite{shapira2002edit}. With the advancement of quantum computing, quantum computers are expected to have the potential to efficiently solve complex problems that are intractable for classical computers \cite{grover1996fast, shor1994algorithms}. Near-term quantum computing on the Noisy Intermediate-Scale Quantum (NISQ) devices \cite{preskill2018quantum} has already shown its ability in many fields, such as quantum chemistry \cite{peruzzo2014variational}, optimization \cite{farhi2014quantum}, and quantum machine learning \cite{havlivcek2019supervised, jerbi2023quantum}. In the task of DNA sequence comparison, quantum computing also holds promise for outperforming classical methods \cite{kosoglu2023biological, varsamis2023quantum}.

We face two main challenges in designing the quantum model: (1) determining how to encode nucleotide bases onto a quantum computer in a principled manner; (2) preserving the pairwise permutation-insensitive property of the EDM metric within the model. Inspired from geometric machine learning \cite{bronstein2021geometric}, incorporating the intrinsic symmetries or invariances of data into a machine learning model is well known to enhance both performance and correctness. Examples include Convolutional Neural Networks (CNNs) \cite{o2015introduction}, which leverage translation invariance in images; Graph Neural Networks (GNNs) \cite{wu2020comprehensive}, which leverage permutation invariance of graph nodes; and recent advances in quantum machine learning \cite{kairon2025equivalence, schatzki2024theoretical, mansky2024permutation}, where symmetries such as permutation invariance are similarly utilized.

In this work, we propose a novel permutation-invariant variational quantum kernel model specifically designed to learn the similarity between DNA sequences defined by EDM. We tackle the above two challenges and contribute as follows:
\begin{itemize}
    \item For challenge (1), we develop a novel encoding circuit for four nucleotide bases to capture the mutual equality among them. 
    \item For challenge (2), we consider DNA sequence similarity defined by EDM and design a permutation-invariant quantum kernel that approximately reflects the pairwise permutation insensitivity property of EDM.  
    \item We conduct numerical experiments to validate the correctness and performance of our method on a quantum simulator and compare to classical deep kernel learning models.
    \item We incorporate the data re-uploading technique \cite{perez2020data, perez2021one} into our model and experimentally investigate how the expressiveness of the quantum kernel improves with increased data re-uploading.
\end{itemize}

Through theoretical analysis and experimental validation, our model has shown its effectiveness for DNA sequence comparison with polynomial circuit depth. Additionally, the broad applicability of kernel functions opens potential directions for future research and application in downstream tasks, such as improving sequence data analysis approaches for global challenges like antimicrobial resistance (AMR). 

This paper is organized as follows: Section \ref{prel} overviews the fundamental background knowledge relevant to this work. Section \ref{method} details the construction of our model, highlighting how it leverages symmetry in the encoding and parameterized circuits. Section \ref{exp} presents and analyzes the results of numerical experiments. Section \ref{concl} discusses the model's performance, limitations, and potential future directions.

\section{Preliminary} \label{prel}
In this section, we provide the basic background information relevant to this work. Subsection \ref{VQC} briefly introduces the concept of variational quantum computing, while Subsection \ref{CDS} discusses the EDM, the metric used as the ground truth in this paper for measuring the similarity between DNA sequences. Subsection \ref{permui} emphasizes the pairwise permutation-insensitive symmetry of EDM.

\subsection{Variational Quantum Computing} \label{VQC}
\begin{figure}[ht]
  \centering
  \includegraphics[width=0.3\textwidth]{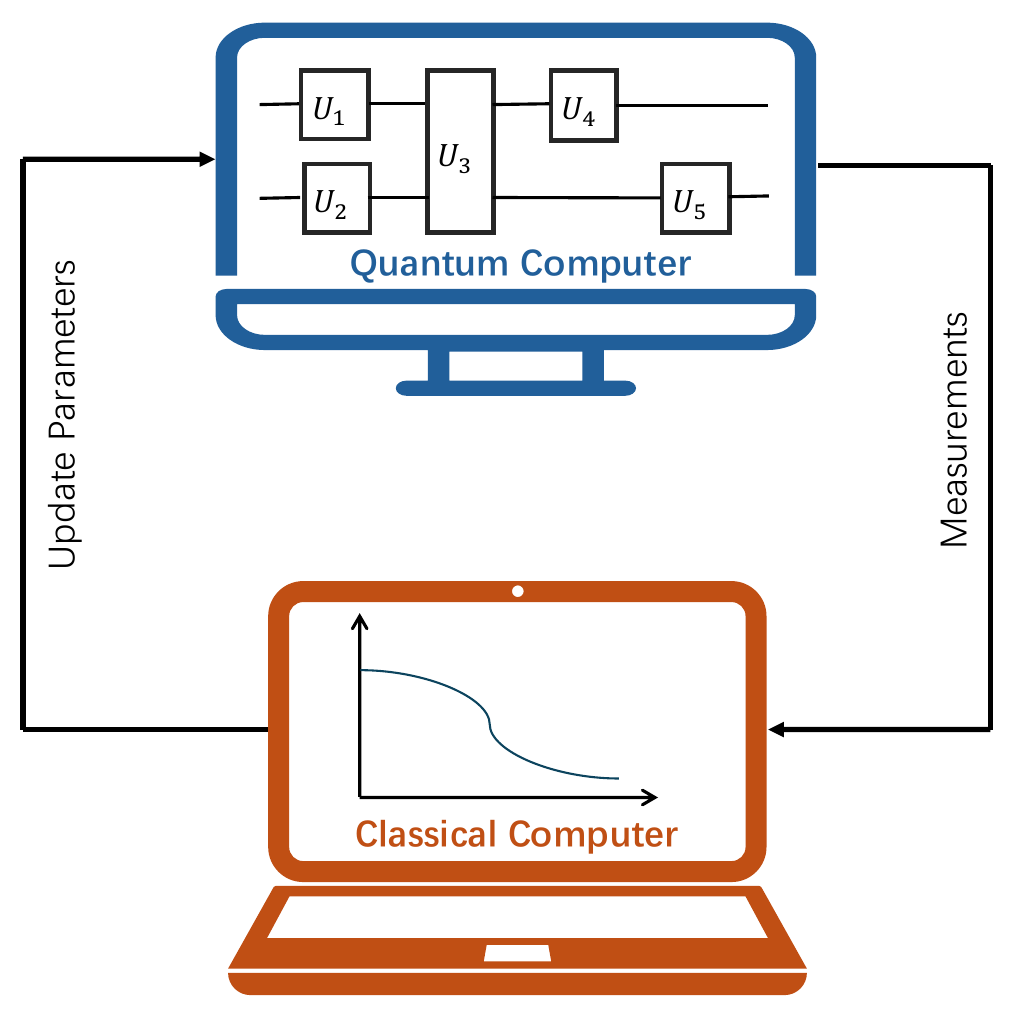}
  \caption{The working paradigm of variational quantum computing. The quantum computer applies a variational quantum circuit to process information or construct models for a certain task. The measurement results will be fed to a classical computer for optimization. The optimizer on the classical computer will update the parameters in the quantum circuit. After multiple loops, the parameters in the quantum circuit will be adjusted to be suitable for solving the target task.  }
  \label{fig:vqc}
\end{figure}

Variational quantum computing is a hybrid quantum-classical approach that leverages the strengths of both quantum and classical computing. In this paradigm, a parameterized quantum circuit (also referred to as a variational quantum circuit) is employed to process information on quantum devices. The measurements from the quantum circuit are then passed to an optimizer on a classical computer for a specific target task, such as minimizing the expectation value of quantum measurements. The classical optimizer analyzes the quantum measurement output and updates the parameters of the quantum circuit accordingly. Through multiple iterations of this loop, the parameters are gradually adjusted to optimize performance for the target task.

Variational quantum computing has attracted significant research attention for the current NISQ devices \cite{preskill2018quantum}. Numerous algorithms based on variational quantum computing have been proposed, with the potential to achieve quantum advantage over classical methods. Notable examples include the Variational Quantum Eigensolver (VQE) \cite{peruzzo2014variational} for quantum chemistry and the Quantum Approximate Optimization Algorithm (QAOA) \cite{farhi2014quantum} for optimization.

Among these algorithms, quantum machine learning has garnered much attention. In quantum machine learning, a quantum circuit consisting of a parameterized circuit $U(\theta)$ and an encoding circuit $V(x)$ is commonly used to construct a machine learning model $f_{\theta}$. The expectation value of quantum measurements with respect to an observable $O$ serves as the model's output, which can be formulated as:
\begin{equation}
f_{\theta}(x)=\bra{0}U^{\dag}(\theta)V^{\dag}(x)OV(x)U(\theta)\ket{0}
\label{qml}
\end{equation}

By feeding this output into a classical optimizer, the parameters of the model $f_{\theta}$ are iteratively updated to solve the target machine learning task. A number of variational quantum algorithms have been developed for machine learning applications, including the Variational Quantum Classifier \cite{farhi2018classification}, the Quantum Expectation Value Sampler \cite{romero2021variational}, and the Variational Quantum Kernel \cite{jerbi2023quantum}. In this work, we focus on the variational quantum kernel and introduce novel features specifically for DNA sequence comparison. The details of our method are discussed in Section \ref{method}.

\subsection{Similarity between DNA Sequences} \label{CDS}
Evaluating the similarity between DNA sequences is a fundamental task in bioinformatics and comparative genomics. The study of sequence comparison algorithms originated with the Needleman-Wunsch algorithm for global sequence alignment \cite{needleman1970general}, which is a generalization of the Levenshtein edit distance \cite{levenshtein1966binary}.

The Levenshtein edit distance between two strings is defined as the minimum number of edits required to transform one string into the other, where the allowed operations include insertion, deletion, and substitution of a single character. Under certain parameter settings, the Needleman-Wunsch algorithm is equivalent to computing the Levenshtein distance between two DNA sequences \cite{sellers1974theory}. Therefore, Levenshtein distance can be applied as a metric to measure the similarity between DNA sequences.

Based on the standard Levenshtein distance, the edit distance with move operations (EDM) has been of interest in biological sequence analysis \cite{shapira2002edit}. In addition to the character-wise operations allowed in standard Levenshtein distance, EDM further permits substring-move operations. However, the exact classical computation of EDM has been proven to be NP-Complete \cite{shapira2002edit}, which poses a significant obstacle to applying EDM for measuring the similarity between DNA sequences. In this work, we use this computationally challenging similarity measure as the ground truth for evaluating the similarity between pairs of DNA sequences.

\subsection{Permutation Insensitivity in EDM} \label{permui}

In this subsection, we emphasize the pairwise permutation insensitivity inherent in EDM. Let $x=[x_1, x_2,...,x_n]$ and $y=[y_1, y_2,...,y_n]$ be two DNA sequences of length $n$, and $D_E(x, y)$ be EDM between $x$ and $y$. The pairwise permutation operator $\Pi_{ij}$ ($1\leq i,j\leq n$) represents the operation of swapping the bases at positions $i$ and $j$. For example, applying $\Pi_{12}$ to $x$ results in $\Pi_{12}(x)=[x_2,x_1,...,x_n]$. It can be verified that EDM changes by at most $2$ when \textbf{the same} pairwise permutation is applied to both input sequences $x$ and $y$:
\begin{align}
    |D_{E}(x,y) - D_{E}(\Pi_{ij}(x),\Pi_{ij}(y))|\leq2
    \label{pivariant}
\end{align}

This kind of property is known as pairwise permutation insensitivity \cite{roy2021adversarial,grainger2023paca}. We provide a simple proof of this conclusion in Equation \ref{pivariant}. Let $L=D_E(x,y)$ while $L'=D_{E}(\Pi_{ij}(x),\Pi_{ij}(y))$. Note that transforming $x$ into $\Pi_{ij}(x)$ requires only one  move operations at most, and the same holds for transforming $\Pi_{ij}(x)$ to $x$. The same reasoning also applies to $y$. If $L>L'$, then we can always transform $x$ and $y$ into $\Pi_{ij}(x)$ and $\Pi_{ij}(y)$ by at most $2$ move operations in total, which indicates that $L-L'\leq2$. Similarly, if $L<L'$, we can derive that $L'-L\leq2$. Therefore, we conclude that Equation~\ref{pivariant} holds. Details in Subsection \ref{paramed}, our model is specifically designed to approximately capture the pairwise permutation-insensitive property of EDM, aiming to achieve potential performance advantages.

\section{Methodology} \label{method}
In this section, we use the most abstract variational quantum kernel model as the starting point and show how to incrementally construct our model from sketch to the final concrete version. The symmetric characteristics in the DNA data comparison task taken advantage of in our model will be discussed in detail.

\subsection{Method Sketch: Variational Quantum Kernel} \label{msketch}
\begin{figure}[ht]
  \centering
  \includegraphics[width=0.8\textwidth]{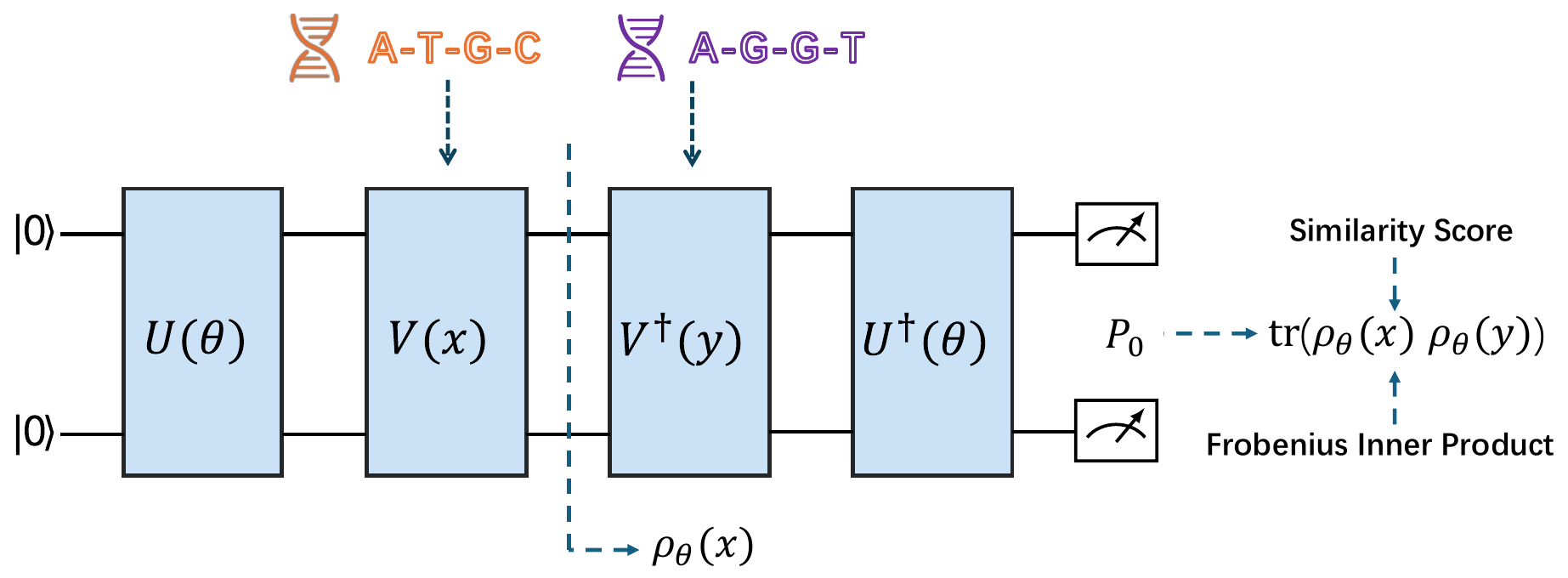}
  \caption{The model sketch of the variational quantum kernel. The model sketch of the variational quantum kernel consists of a parameterized layer $U(\theta)$, an encoding layer $V(\cdot)$, and their corresponding conjugate transpose layers, arranged sequentially. The initial state is set to be $\ket{0}$. The output of the model is the probability of measuring $0$ in the computational basis, which is the Frobenius inner product between two density matrices. Thus, the entire model constructs a kernel function to metric similarity using a variational quantum circuit.}
  \label{fig:modelsketch}
\end{figure}

Our model sketch begins with the most abstract representation of a variational quantum kernel, as illustrated in Figure \ref{fig:modelsketch}. The variational quantum kernel is a specialized form of a variational quantum circuit, consisting of a parameterized layer $U(\theta)$, an encoding layer $V(\cdot)$, and their corresponding conjugate transpose layers $V^{\dag}(\cdot)$ and $U^{\dag}(\theta)$. The initial state is the zero state $\ket{0}$. The probability of measuring $0$ is used as the output of the quantum model. 

For a pair of inputs $x$ and $y$, the output of the quantum model $K_{\theta}$ is given by:
\begin{align}
    K_{\theta}(x,y) &= \vert\bra{0}U^{\dag}(\theta)V^{\dag}(y)V(x)U(\theta)\ket{0} \vert^2\notag \\
    &= \bra{0}U^{\dag}(\theta)V^{\dag}(x)V(y)U(\theta)\ket{0}\bra{0}U^{\dag}(\theta)V^{\dag}(y)V(x)U(\theta)\ket{0} \notag \\
    &= \tr\big(\rho_{\theta}(x)\rho_{\theta}(y)\big) 
\end{align}
where the first equality follows from the Born rule \cite{nielsen2001quantum}, while the third equality uses the density operator representation of quantum states. Specifically, $\rho_{\theta}(x)=V(x)U(\theta)\ket{0}\bra{0}U^{\dag}(\theta)V^{\dag}(x)$ and $\rho_{\theta}(y)=V(y)U(\theta)\ket{0}\bra{0}U^{\dag}(\theta)V^{\dag}(y)$. 

Note that $\rho_\theta(\cdot)$ can be viewed as a parameterized feature mapping function from the input space $\mathcal{X}$ to the feature space. According to the definition of the Frobenius inner product, the output of the model $K_{\theta}(x,y)$ represents the inner product between the density matrices $\rho_{\theta}(x)$ and $\rho_{\theta}(y)$. Therefore, by the definition of the kernel method \cite{shawe2004kernel}, the quantum model $K_{\theta}:\mathcal{X}\times \mathcal{X} \rightarrow \mathbb{R}$ qualifies as a kernel function.

In our case, the inputs $x$ and $y$ represent two DNA sequences to be compared. The quantum kernel $K_{\theta}$ evaluates the similarity between $x$ and $y$ and outputs a score $K_{\theta}(x,y)$ between $0$ and $1$. The more similar the two sequences are, the higher the score will be. Notably, if the two input DNA sequences are identical, the output score will be $1$.

In the following subsections, we will construct this abstract model step by step, methodically building from the foundation to the final concrete version tailored to our task.
\subsection{Encoding Layer: Mutual Equality} \label{encoding}
\begin{figure}[ht]
    \centering
    \begin{minipage}{0.45\textwidth}
        \centering
        \begin{tabular}{|c|c|}
            \hline
            Nucleotide Bases & Quantum state representation \\
            \hline 
            Adenine (A) & $\ket{0}$ \\
            Thymine (T) & $\frac{1}{\sqrt{3}}\ket{0}+\sqrt{\frac{2}{3}}\ket{1}$ \\
            Guanine (G) & ${\frac{1}{\sqrt{3}}}\ket{0}+\sqrt{\frac{2}{3}}e^{i\frac{2\pi}{3}}\ket{1}$ \\
            Cytosine (C) & ${\frac{1}{\sqrt{3}}}\ket{0}+\sqrt{\frac{2}{3}}e^{i\frac{4\pi}{3}}\ket{1}$ \\
            \hline
        \end{tabular}
        \captionof{table}{In the encoding layer, each nucleotide base is assigned a unique quantum state on its corresponding qubit. These four quantum states are known as the SIC-POVM states.}
        \label{tab:angles}
    \end{minipage}
    \hfill
    \begin{minipage}{0.45\textwidth}
        \centering
        \includegraphics[width=\linewidth]{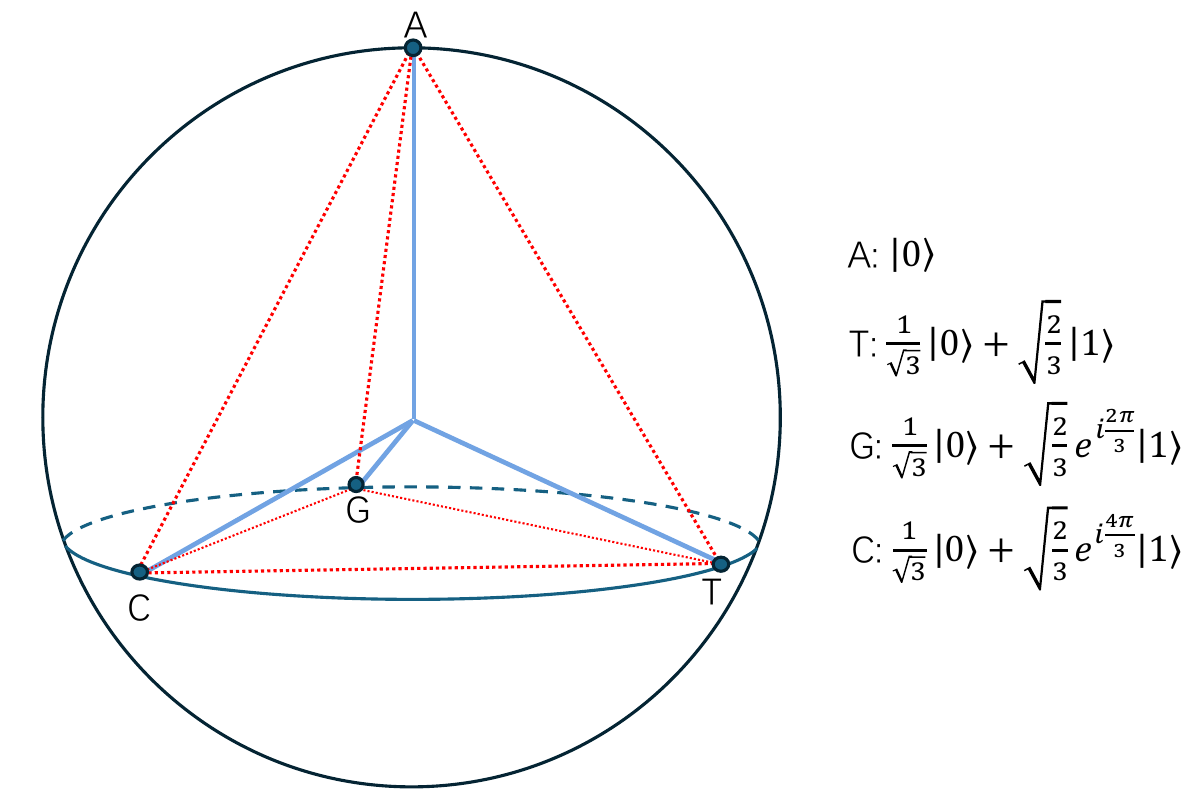} 
        \captionof{figure}{The four SIC-POVM states form a regular tetrahedron on the Bloch sphere to capture the mutual equality for encoding.}
        \label{fig:sicp}
    \end{minipage}
\end{figure}

In this subsection, we will construct the encoding layer $V(\cdot)$ for our task. A DNA sequence can be represented as a string, where each position corresponds to one of the four nucleotide bases: A, T, G, or C. Our overall strategy of encoding is to encode each base using a single qubit and stack these qubits together to represent the entire DNA sequence. For each base, we map it to a unique single qubit. The specific quantum states associated with each base are shown in \Cref{tab:angles}. These states are known as the SIC-POVM states for a single-qubit quantum system \cite{renes2004symmetric}. 

An important feature of these states is their symmetric property, which allows us to capture the mutual equality when encoding the four bases. Simply put, without biases of the parameterized layers, the similarity between A and G should not be greater than the similarity between A and C. Therefore, in each qubit, the encoding layer must treat the four bases equally, ensuring that in the encoding space, the distance between any pair of the quantum states representing the four bases is identical.

It is evident that the SIC-POVM states fulfill this requirement. As shown in \Cref{fig:sicp}, the four states form a regular tetrahedron on the Bloch sphere, with each vertex representing a state (corresponding to a base). This highly symmetric property ensures that our encoding layer maintains the mutual equality among the four bases on each qubit.

This method also takes advantage of the biological characteristics of DNA sequences. Notably, there can be at most four states on the Bloch sphere that satisfy the condition of mutual equality. This can be understood geometrically: in three-dimensional space, there are at most four points where the distance between any two points is equal. For DNA sequences, there are exactly four bases, which makes our strategy of encoding effective. 

\begin{figure}[ht]
  \centering
  \includegraphics[width=0.3\textwidth]{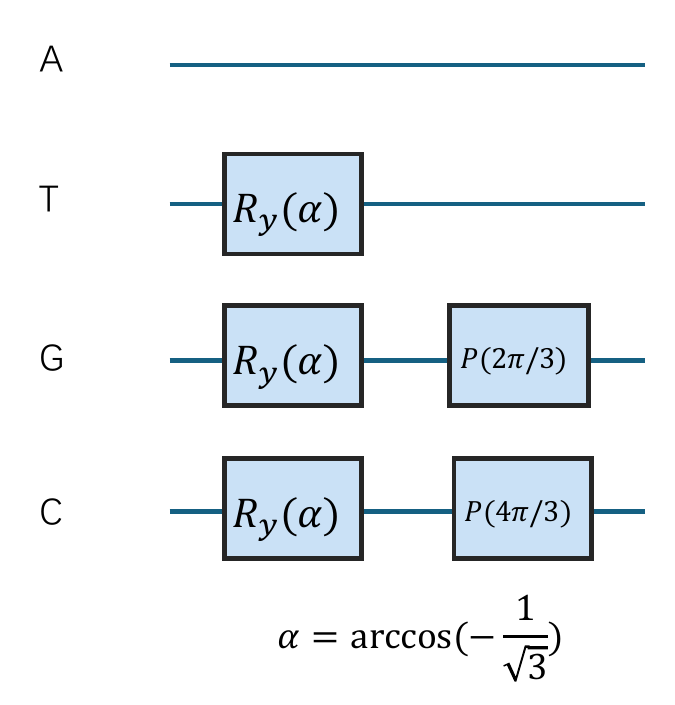}
  \caption{The encoding layer for the length-4 DNA sequence ``ATGC''. For Adenine, there is no quantum gate because only an identity operator is needed. For the other bases, a rotation gate $R_y$ and a phase gate $P$ are applied with corresponding angles.}
  \label{fig:egencode}
\end{figure}

In \Cref{fig:egencode}, we provide an illustrative example of a quantum circuit used to encode the length-4 DNA sequence ``ATGC'' following our encoding strategy. This example also shows how to encode the four bases to the corresponding quantum states using quantum gates. 

\subsection{Parameterized Layer: Permutation Invariance} \label{paramed}
In this subsection, we construct the encoding layer $U(\theta)$ for our task. As mentioned in \Cref{CDS}, the similarities between two DNA sequences can be quantified using the edit distance. In our case, we primarily consider EDM $D_E$ as the metric.

As discussed in \Cref{permui}, EDM exhibits a pairwise permutation-insensitive symmetry. A direct implication is that a well-designed kernel for evaluating the similarity between two DNA sequences should exhibit the same symmetry. However, due to the difficulty of directly incorporating permutation insensitivity into the model, we use permutation invariance to approximately capture the desired permutation-insensitive symmetry. Therefore, the variational quantum kernel $K_{\theta}$ used in our task is expected to satisfy the pairwise permutation-invariant property:
\begin{align}
    K_{\theta}(x,y) = K_{\theta}(\Pi_{ij}(x),\Pi_{ij}(y))
    \label{kivariant}
\end{align}

Permutation invariance is not a universal property of variational quantum circuits or quantum kernels. Therefore, to ensure that the quantum kernel exhibits this property, we must carefully design the quantum circuit.

Since there are no entanglement gates in the encoding layer, we can verify that if the parameterized circuit $U_{\theta}$ is permutation-invariant, then the quantum kernel will satisfy \Cref{kivariant}. Following the definition in \cite{mansky2024permutation}, a quantum circuit $U$ is said to be permutation-invariant if it satisfies the condition: 
\begin{equation}
    U = U_{\Pi_{ij}} UU_{\Pi_{ij}}
    \label{ivc}
\end{equation}
where $ U_{\Pi_{ij}}$ represents the $\operatorname{SWAP}$ gate acting on the $i$-th and $j$-th qubits. Furthermore, since there are no entanglement gates in the encoding layer, it is straightforward to verify that:
\begin{equation}
    V(\Pi_{ij}(x)) =  U_{\Pi_{ij}}V(x)U_{\Pi_{ij}}
    \label{ien}
\end{equation}

Suppose the parameterized layer is permutation-invariant, following the \Cref{ivc} and \Cref{ien}, we have:
\begin{align}
    K_{\theta}(\Pi_{ij}(x),\Pi_{ij}(y)) &= \bra{0}U^{\dag}(\theta)V^{\dag}(\Pi_{ij}(x))V(\Pi_{ij}(y))U(\theta)\ket{0}\bra{0}U^{\dag}(\theta)V^{\dag}(\Pi_{ij}(y))V(\Pi_{ij}(x))U(\theta)\ket{0} \notag \\
    &= \bra{0}U^{\dag}(\theta)U_{\Pi_{ij}}V^{\dag}(x)U_{\Pi_{ij}}U_{\Pi_{ij}}V(y)U_{\Pi_{ij}}U(\theta)\ket{0}\bra{0}U^{\dag}(\theta)U_{\Pi_{ij}}V^{\dag}(y)U_{\Pi_{ij}}U_{\Pi_{ij}}V(x)U_{\Pi_{ij}}U(\theta)\ket{0} \notag \\
    &= \bra{0}U_{\Pi_{ij}}U^{\dag}(\theta)V^{\dag}(x)V(y)U(\theta)U_{\Pi_{ij}}\ket{0}\bra{0}U_{\Pi_{ij}}U^{\dag}(\theta)V^{\dag}(y)V(x)U(\theta)U_{\Pi_{ij}}\ket{0} \notag \\
    &= \bra{0}U^{\dag}(\theta)V^{\dag}(x)V(y)U(\theta)\ket{0}\bra{0}U^{\dag}(\theta)V^{\dag}(y)V(x)U(\theta)\ket{0} \notag \\
    &= K_{\theta}(x,y)
\end{align}

Thus, we have proven that if the parameterized layer is permutation-invariant and the encoding layer follows the setting in \Cref{encoding}, then the quantum kernel $K_{\theta}$ satisfies \Cref{kivariant}, ensuring permutation invariance.

\begin{figure}[ht]
  \centering
  \includegraphics[width=0.3\textwidth]{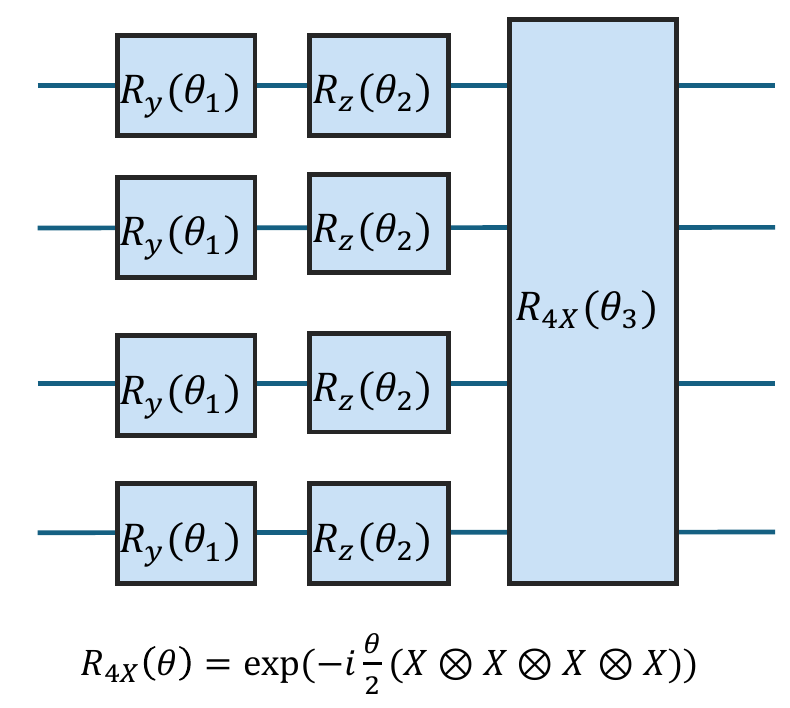}
  \caption{The parameterized layer for the length-4 DNA sequence. The $R_{NX}$ gate is applied to provide entanglement, after a series of $R_z$ and $R_y$ gates. The parameterized layer provides three trainable parameters. This parameterized layer is permutation-invariant, which can make the quantum kernel fulfill the permutation invariance property in Formula \ref{kivariant}.}
  \label{fig:parametrized}
\end{figure}

Following the design principles for permutation-invariant circuits outlined in \cite{mansky2024permutation}, we use the following simple quantum circuit as the parameterized layer. A 4-qubit example is illustrated in \Cref{fig:parametrized}. The circuit consists of an $R_{NX}$ gate to introduce entanglement. The $R_{NX}$ gate can be represented as $R_{NX}(\theta)=\operatorname{exp}\big({-i\frac{\theta}{2}(X\otimes \overset{n}{\cdots} \otimes X)}\big)$. Then, a single qubit rotation gate $R_z$ and $R_y$ are applied. Specifically, the parameterized layer provides three trainable parameters. One can verify that this circuit satisfies the permutation-invariant property defined by \Cref{kivariant}.

\subsection{Data Re-uploading: Expressiveness} \label{drea}
\begin{figure}[ht]
  \centering
  \includegraphics[width=0.9\textwidth]{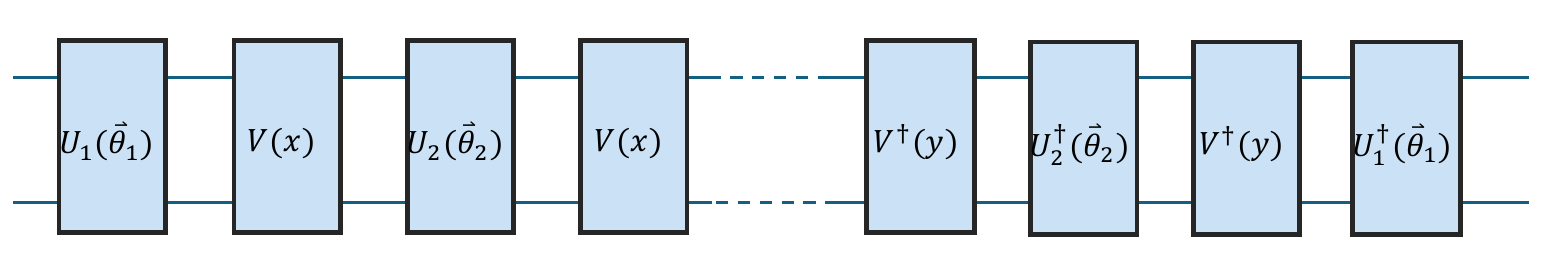}
  \caption{The data re-uploading technique alternates between parameterized layers and encoding layers. It has been proven to improve the model's expressive capability. Notably, the conjugate transpose components also appear alternately in the correct order, ensuring that the permutation-invariant property is preserved.}
  \label{fig:dr}
\end{figure}

Instead of using only a single parameterized layer and one encoding layer with their conjugate transposes, we can incorporate multiple layers to enhance the model's expressive capability, following the data re-uploading technique \cite{perez2020data, perez2021one}. As shown in \Cref{fig:dr}, the data re-uploading technique alternates between parameterized layers and encoding layers, which is proven to be able to improve the model's expressiveness. Besides, one can verify that applying the data re-uploading technique does not break the permutation-invariant property of the quantum kernel.

\subsection{The Model: Put It All Together} \label{analysis}
By incorporating all the specific details discussed above into the abstract sketch from \Cref{msketch}, we obtain our final model. First, the model employs the encoding layer from \Cref{encoding} to capture the mutual equality among nucleotide bases. Next, it utilizes a permutation-invariant quantum circuit as the parameterized layer. It ensures that the entire quantum kernel maintains permutation invariance, which approximately captures the essential permutation insensitivity property for measuring the similarity between DNA sequences. Finally, the data re-uploading technique is applied to enhance the model's expressiveness. 

On a NISQ device, quantum circuits are generally limited to shallow depths that grow no faster than polynomially with the system size $n$. In our model, $n$ corresponds to the length of the DNA sequences. Therefore, the computational complexity of evaluating our quantum kernel—i.e., computing a single approximate similarity score between two DNA sequences—remains polynomial to the sequence length $n$.

\section{Numerical Experiments} \label{exp}
In this section, we evaluate the performance of the permutation-invariant quantum kernel in comparing the similarity between DNA sequences by numerical experiments on classical simulators. The experimental setup is detailed in \Cref{ess}, and the results are presented in \Cref{results}. We also provide a brief introduction to the classical kernel learning models used for comparison in \Cref{clas}.

\subsection{Experiment Setting} \label{ess}
Due to the limitations of current quantum devices and classical simulators, our numerical experiments are restricted to comparing short DNA sequences, each consisting of eight nucleotide bases. All experimental results are obtained through classical simulations.

As discussed in \Cref{analysis}, since the DNA sequences have a length of $8$, the quantum kernel in our experiments utilizes eight qubits. To enhance the model's expressive capability, the number of data re-uploading layers is set to $24$, resulting in $72$ trainable parameters in the quantum kernel model. In order to evaluate the effect of data re-uploading
technology for improving the model’s expressiveness, we also chose the number of data re-uploading as $6$ and $12$ to run the experiments for a comparison to $24$. In the following, we denote our method as QKernel-$N$, where $N$ represents the number of data re-uploading layers.

For many tasks in bioscience, relative similarity scores are sufficient in practical applications. For example, when searching for the most similar sequences in a gene sequence repository for a given query DNA sequence $A$, it is only necessary to identify which sequence has the highest similarity with $A$. In this case, the absolute value of the similarity score is less important, whereas the relative ranking of similarity scores is crucial. Inspired by this fact, we propose using \textbf{order accuracy} as the metric to evaluate the model's performance.

Order accuracy evaluates whether the model preserves the similarity score rankings between two pairs of sequences. Specifically, for three DNA sequences $A$, $B$ and $C$, if the ground truth similarity score (EDM $D_E$ in this paper) shows the relation $D_E(A,B)>D_E(A,C)$, then a well-performing model should maintain the relative order, meaning that $K_{\theta}(A,B)<K_{\theta}(A,C)$. Note that similarity scores and the distances are inversely related. Order accuracy quantifies the percentage of DNA sequence triplets for which the quantum kernel model correctly preserves the relative similarity rankings. 

The training and test set each contain $3200$ triplets of DNA sequences, totaling $9600$ sequences per set. The ground truth for training is based on EDM, which is transformed and normalized into a similarity score between $0$ and $1$ as follows:
\begin{equation}
  S_T(A,B)=\frac{N-D_E(A,B)}{N}
  \label{simi}
\end{equation}
where $N$ is the length of DNA sequences (namely 8 in our experiments), and $D_E(A,B)$ denotes EDM between DNA sequences $A$ and $B$. $S_T$ serves as the ground truth similarity score in our experiments.

The optimizer used for training is the commonly used stochastic gradient descent (SGD) \cite{ruder2016overview} with a learning rate of $0.01$. The loss function is the mean square error (MSE) between the ground truth and the model's prediction. The model is trained for $100$ epochs, and its performance is evaluated on the test set after each epoch. 

\subsection{Classical Deep Kernel Learning Models for Comparison} \label{clas}
\begin{figure}[ht]
    \centering
    \includegraphics[width=0.6\textwidth]{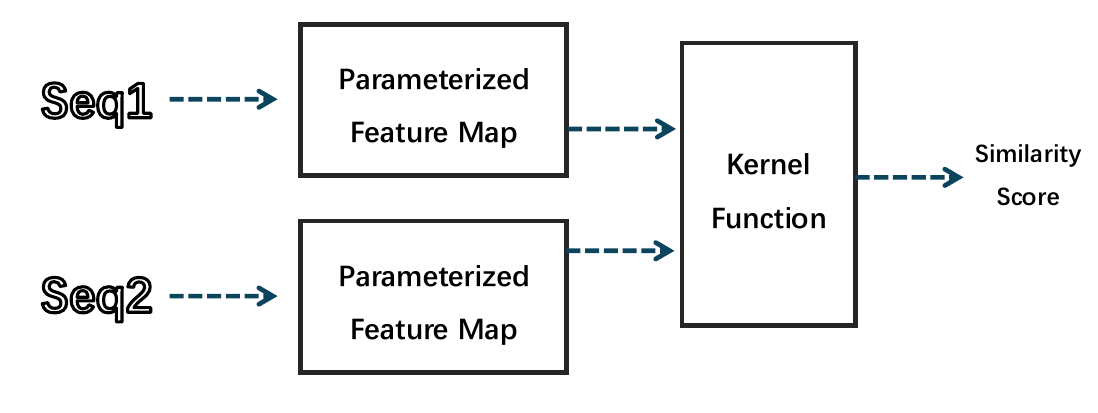} 
    \captionof{figure}{The classical deep kernel learning models used for comparison. Each pair of sequences is embedded through a parameterized feature map, which is detailed in \Cref{tab:classical}. The resulting embeddings are then fed into a kernel function to produce the final similarity score. For the kernel functions, we use RBF, cosine and poly2 for comparision.}
    \label{fig:classical}
\end{figure}

The quantum kernel model we propose can be regarded as a type of kernel learning model \cite{wilson2016deep}. To assess the efficiency of our approach, we compare it with classical deep kernel learning models. To ensure a fair comparison to the greatest extent possible, we restrict the classical models to simple architectures (as illustrated in \Cref{fig:classical}), even though they already contain far more trainable parameters than our quantum kernel model (as reported in \Cref{tab:classical}).

The classical kernel learning model used for comparison consists of a parameterized feature map that embeds each DNA sequence into a vector representation. A kernel function is then applied to these embeddings to produce a similarity score. In our experiments, we consider three kernel functions: RBF, cosine, and poly2.

The structure of the parameterized feature map used in the comparison experiments is shown in \Cref{tab:classical}. First, the trainable embedding layer maps each of the four nucleotides to a $4$-dimensional vector. Hence, a length-$8$ DNA sequence is represented as an $(8,4)$-tensor after embedding. This tensor is then flattened into a $32$-dimensional vector, which serves as the input to a two-layer MLP. Although this parameterized feature map has a simple architecture, it already contains far more trainable parameters than our quantum kernel model.

\begin{table}[ht]
\centering
\begin{tabular}{ccc}
\toprule
    \textbf{Layers} & \textbf{Output Size} & \textbf{Number of Trainable Parameters}\\ 
    \midrule
    Embedding & $(8,4)$  & $16$ \\
    Flatten & $(32,)$ & $0$ \\
    Linear & $(16, )$ & $528$ \\
    ReLU & $(16, )$ & $0$ \\
    Linear & $(16, )$ & $272$ \\
    \bottomrule &
\end{tabular}
\captionof{table}{The parameterized feature map used in the classical deep kernel learning models shown in \Cref{fig:classical}. The settings are for length-$8$ DNA sequences, which contain $816$ trainable parameters in total.}
\label{tab:classical}
\end{table}

\subsection{Results} \label{results}
\begin{figure}[ht]
  \centering
  \includegraphics[width=0.6\textwidth]{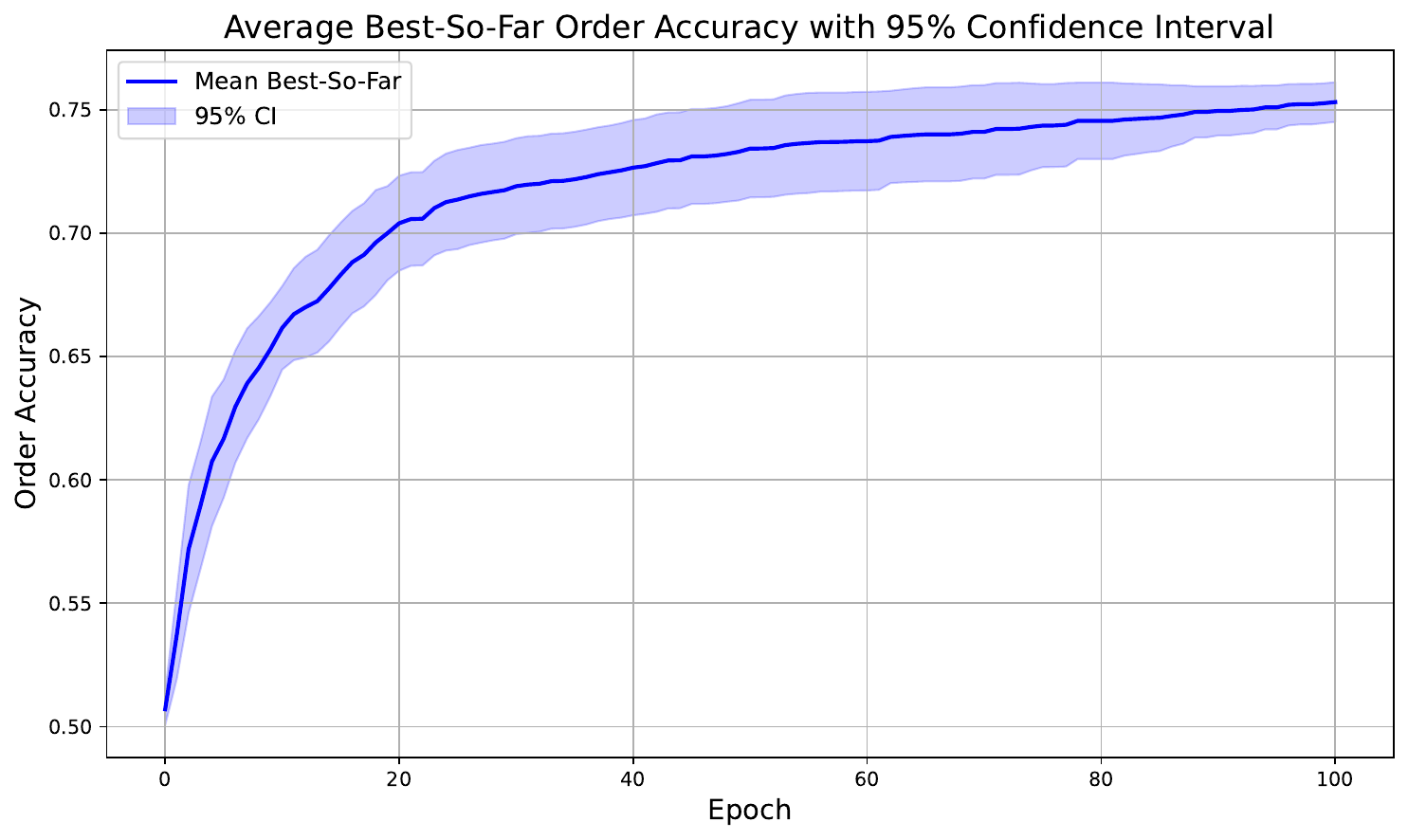}
  \caption{Best-so-far learning curves with confidence interval. As the metric, the order accuracy is evaluated at the end of each training epoch. The model learns quickly at the beginning and improves at a slower rate in the later epochs. After $100$ epochs of training, the model achieves an order accuracy of over $75\%$, indicating that it effectively learns during training.}
  \label{fig:result1}
\end{figure}

The training and evaluation process is conducted independently $10$ times to mitigate the impact of randomness. For QKernel-$24$, the average best-so-far learning curve along with the $95\%$ confidence interval on the test set, is shown in \Cref{fig:result1}. We observe that the average order accuracy increases throughout the training process. The model improves rapidly at the beginning of training, followed by a gradual improvement at a slower rate.

The Qkernel-$24$ model begins with random guessing. For order accuracy, a completely random guess would result in $50\%$ accuracy. After $100$ epochs of training, the model achieves an average order accuracy of over $75\%$, which indicates that it successfully learns the characteristics of similarity between DNA sequences during training.

For quantum kernel models with different numbers of data re-uploading and for the classical deep kernel learning models, we use the same training and evaluation settings. The averaged best order accuracies achieved over 10 runs, together with the total number of trainable parameters in each model, are reported in \Cref{tab:datarp}.

\begin{table}[ht]
    \centering
    \begin{tabular}{ccc}
        \toprule
        \textbf{Kernel Learning Model} & \textbf{Order Accuracy} & \textbf{Number of Trainable Parameters}\\ 
        \midrule
        QKernel-$24$ & $75.3\pm1.3\%$  & $72$\\
        Qkernel-$12$ & $73.8\pm 3.2\%$ & $36$\\
        Qkernel-$6$ & $66.7\pm 2.9\%$ & $18$\\
        Ckernel-RBF & $59.1\pm1.3\%$ &  $817$\\
        Ckernel-cosine & $58.9\pm1.1\%$ &  $816$\\
        Ckernel-poly2 & $59.7\pm0.77\%$ &  $818$\\
        \bottomrule &
    \end{tabular}
    \caption{The average best order accuracy achieved over $10$ runs, corresponds to different data re-uploading numbers and different classical kernel learning models. We observe that order accuracy increases as the number of data re-uploading increases, indicating that the data re-uploading technique effectively enhances the expressiveness of the quantum kernel model. Besides, although classical kernel learning model contains more trainable parameters, its performance is far worse compared to the quantum kernel.}
    \label{tab:datarp}
\end{table}

For quantum kernels, we observe that the order accuracy increases as the number of data re-uploading layers grows, indicating that the data re-uploading technique effectively enhances the expressive capacity of the model. Therefore, our quantum kernel is expected to achieve better performance with more data re-uploading layers.

For classical deep kernel learning models, we find that they perform significantly worse than the quantum kernel, even though they contain far more trainable parameters. The averaged order accuracies of all classical models remain around $60\%$. These results indicate that the specially designed permutation-invariant quantum kernel with mutually equivalent encoding substantially improves the efficiency of kernel learning. With far fewer parameters, our quantum kernel achieves considerably more promising results.

\section{Discussion and Conclusion} \label{concl}
In this work, we construct a permutation-invariant quantum kernel designed to compare the similarity between DNA sequences while leveraging the intrinsic pairwise permutation-insensitive symmetry of EDM. We develop an encoding circuit tailored for nucleotide bases, ensuring that it captures their mutual equality. To realize permutation invariance in the quantum kernel, we incorporate a permutation-invariant circuit. Additionally, we employ the data re-uploading technique to enhance the model's expressivity. Numerical experiments indicate that despite utilizing only a few trainable parameters, the quantum kernel effectively measures the relative similarity between DNA sequences.

EDM is NP-Complete to compute exactly in the general case. Our quantum kernel model can be viewed as a surrogate model that approximates the similarity measure defined by EDM while allowing polynomial-time inference. It is worth noting that this comparison is not entirely fair, since the quantum kernel model requires a training phase before inference. Therefore, we additionally compare our method with several classical deep kernel learning models under the same training budget. Our quantum kernel consistently achieves superior order accuracy, even though the classical models employ far more trainable parameters. This suggests that the specially designed symmetries incorporated into our quantum kernel architecture substantially improve its efficiency as a surrogate model for the similarity measure defined by EDM.

Our model still has certain limitations due to the constraints of current quantum devices. The scalability of quantum devices is hindered by factors such as noise \cite{preskill2018quantum} and the barren plateaus problem \cite{mcclean2018barren}, making it currently almost impossible to process real DNA sequences directly. This is primarily due to the circuit width required to represent the sequences. In our work, we only evaluate the model on DNA sequences of length $8$. Although the model should theoretically remain effective for longer sequences, its performance has yet to be experimentally verified on large-scale data. 

The scalability limitations of NISQ devices also constrain the circuit depth. In our work, we employ a relatively simple permutation-invariant parameterized circuit along with a limited number of data re-uploading iterations. While this simplification improves practicality on current hardware, it inevitably limits the model's expressive capacity. As a result, although the absolute order accuracy achieved by our method is not yet high, it still surpasses that of classical kernel learning models with larger parameter counts. Our experiments further indicate that increasing the number of data re-uploading layers enhances the performance of the model. Also, our method is expected to benefit from more sophisticated permutation-invariant circuit architectures for higher expressivity. General principles for constructing such expressive permutation-invariant quantum circuits can be found in \cite{mansky2024permutation}.

In this work, we use permutation invariance in the design of the quantum circuit to approximately capture the permutation insensitivity inherent in EDM between DNA sequences. However, this approximation may introduce error, as the desired symmetry is not exactly preserved. Designing a permutation-insensitive quantum kernel model directly instead of permutation invariance is more complex, but remains a meaningful direction for future work.

Like other general kernel methods, our quantum kernel model is theoretically applicable to a variety of downstream tasks in bioscience. The most direct application is identifying genes of interest (i.e., those similar to a reference DNA sequence) in gene sequence data repositories, including antimicrobial resistance (AMR) gene detection \cite{feldgarden2021amrfinderplus, galhano2021antimicrobial}, novel gene identification \cite{klasberg2016computational}, and homology search \cite{reeck1987homology}. Additionally, as a kernel function, our model naturally integrates with support vector machines (SVMs) to build classification models for bioscience applications, such as AMR gene classification \cite{feldgarden2021amrfinderplus}. Validation of the model on downstream tasks is beyond the scope of this paper. However, we consider these applications as potential directions for future work. As quantum devices become more powerful, our model is expected to handle more complex cases involving longer DNA sequences.

\bibliographystyle{unsrt}  
\bibliography{references}
\end{document}